\documentclass[12pt,onecolumn]{article}

\usepackage{scicite}

\usepackage[dvips]{graphicx}
\usepackage{times}

\newenvironment{sciabstract}{%
\begin{quote} \bf}
{\end{quote}}
\newcommand\araa{Annu. Rev. Astron. Astrophys.}
\newcommand\apjl{Astrophys. J.}
\newcommand\aj{Astron. J.}
\newcommand\aap{Astron. Astrophys.}

\newcommand\aaps{Astron. Astrophys. Suppl. Ser.}
\newcommand\apj{Astrophys. J.}
\newcommand\mnras{Mon. Not. R. Astron. Soc.}
\newcommand\apjs{Astrophys. J. Suppl. Ser.}
\newcommand\bain{Bull. Astron. Inst. Netherlands}

\newcommand\aaspb{AAS Photo-Bull.}

\newcounter{lastnote}
\newenvironment{scilastnote}{%
\setcounter{lastnote}{\value{enumiv}}%
\addtocounter{lastnote}{+1}%
\begin{list}%
{\arabic{lastnote}.}
{\setlength{\leftmargin}{.22in}}
{\setlength{\labelsep}{.5em}}}
{\end{list}}

\title{The Spiral Structure of the \\Outer Milky Way in Hydrogen}

\author
{E.S. Levine,$^{\ast}$ Leo Blitz, Carl Heiles\\
\\
\normalsize{Astronomy Department, University of California,}\\
\normalsize{601 Campbell Hall,  Berkeley, CA 94720}\\
\\
\normalsize{$^\ast$To whom correspondence should be addressed; E-mail:  elevine@astron.berkeley.edu.}\\
}

\date{}

\begin{document} 


\baselineskip14pt 


\maketitle


\begin{sciabstract}
We produce a detailed map of the perturbed surface density of neutral hydrogen in the outer Milky Way  disk  demonstrating that the Galaxy is a non-axisymmetric multi-armed spiral. Spiral structure in the southern half of the Galaxy can be traced out to at least 25 kpc, implying a minimum radius for the gas disk. Overdensities in the surface density are coincident with regions of reduced gas thickness. The ratio of the surface density to the local median surface density  is relatively constant along an arm.  Logarithmic spirals can be fit to the arms with pitch angles of 20$^\circ$-25$^\circ$. \end{sciabstract}

\footnotetext{\it Science\rm, in press. Embargoed for discussion in the popular press until publication by \it ScienceXpress\rm.}


Mapping the Milky Way's spiral structure is traditionally difficult  because the Sun is imbedded in the Galactic disk; absorption by dust renders optical methods ineffective at  distances larger than a few kpc. Radio lines like the 21 cm hyperfine transition of atomic hydrogen (HI) are not affected by this absorption, and are therefore well-suited to looking through the disk. The density of HI is roughly proportional to the intensity of the emission, barring optical depth effects. Maps are constructed by using the Doppler shift of the emission line in combination with the rotation structure to determine where in the Galaxy the emission originates. 
The first maps of the HI density in the midplane of the Milky Way \cite{VMO1954} made using the 21 cm  transition offered the promise that HI mapping of the Galactic plane would reveal its spiral structure by highlighting regions with HI overdensities. 
The task is easiest outside of the Sun's orbit around the Galactic center because velocities in this region map into distances uniquely; in the inner Galaxy, however, each velocity corresponds to two distinct distances from the Sun, so an unambiguous velocity-to-position mapping is impossible.
 When observations of the southern plane of the Galaxy made in Australia were combined with northern sky data from the Netherlands  \cite{KW1965}, a picture of spiral structure of the inner Milky Way was seen, but it did not clearly resemble that of a typical spiral galaxy; the spiral structure in the outer Galaxy (outside the Solar orbit) was less apparent. More recent maps of HI in the outer Galaxy have improved on the early work \cite{HJK1982}.

In this paper we present a map of perturbations on the HI surface density in the outer Milky Way  using a modified unsharp masking technique. By subtracting a blurred copy from the original image, this technique emphasizes low contrast features in both bright and dim regions \cite{M1977} such as spiral structure. In previous maps (e.g.~\cite{HJK1982} and Fig.~1) spiral arms were difficult to pick out on top of the surface density's global falloff  with radius.

We construct a grid of the HI  density  for the outer Galactic disk, $\rho$, as a function of Galactocentric radius $R$, angle $\phi$, and height off the plane $z$ from the Leiden/Argentine/Bonn Galactic HI survey \cite{LAB,HB1997,BALMPK2005,ABLMP2000}.
The density is recovered from the measured intensity by assuming a constant spin temperature of 155 K and using the Doppler shift velocity to determine distance along the line of sight. We apply a median filter to remove small features not associated with the disk (such as other galaxies), and a thickness filter to remove larger-scale features like clouds and spurs coming off the disk. We adopt the International Astronomical Union standard value of $R_0=8.5$ kpc for the Sun-Galactic center distance; recent measurements suggesting that this value may be too high would result in a proportional rescaling of the disk \cite{R1993,OM1998}.
 We construct the surface density map, $\Sigma(R,\phi)$,  by summing the grid $\rho(R,\phi,z)$ over $z$ (Fig.~1).
For our rotation structure, we use circular orbits  with a frequency $m=1$ epicyclic streamline correction and a circular velocity of 220 km/s constant with radius \cite{LBH2006a}. 
 The epicyclic correction is necessary because previous studies of HI spiral arms that use purely circular orbits result in grossly discontinuous surface density contours near the Sun-Galactic center line (e.g. \cite{HJK1982}); the magnitude of the correction is fit to minimize this discontinuity. The maximum line-of-sight velocities of the corrections are fixed to lie along the Sun-Galactic center line. 
 We exclude from the map regions within $\pm 15^\circ$ of the Sun-Galactic center line    because of the difficulty of determining reliable distances in these directions. Our radial grid runs over the Galactocentric range 8.6 kpc to 40 kpc.

 For each point on our grid, we calculate $\eta(R,\phi)$, the median surface density of the points within $25^\circ$ in $\phi$ and 2 kpc in $R$; this local median is analogous to the blurred image used in unsharp masking. It is well suited for looking for perturbations in surface density  because it takes into account the falloff in surface density with radius, and adjusts if  the Galaxy  has a lopsided or lumpy distribution.  For points within 2 kpc of the inner and outer radial borders of our grid, we narrow the radial range so that the median is evenly balanced with surface densities inside and outside the point. We then compare the surface density at each point to the median value:
\begin{equation}
\Pi(R,\phi)=\Sigma(R,\phi)/\eta(R,\phi).
\end{equation} 
This is a dimensionless quantity, and therefore is a direct measure of the strength of the surface density perturbations  (Fig.~1). Dividing by the local median rather than subtracting, as is normally done in unsharp masking,  compensates for the change in the surface density by more than an order of magnitude over the radial range of the map. 
 
 \begin{figure}
\begin{center}
\includegraphics[angle=0,scale=.65]{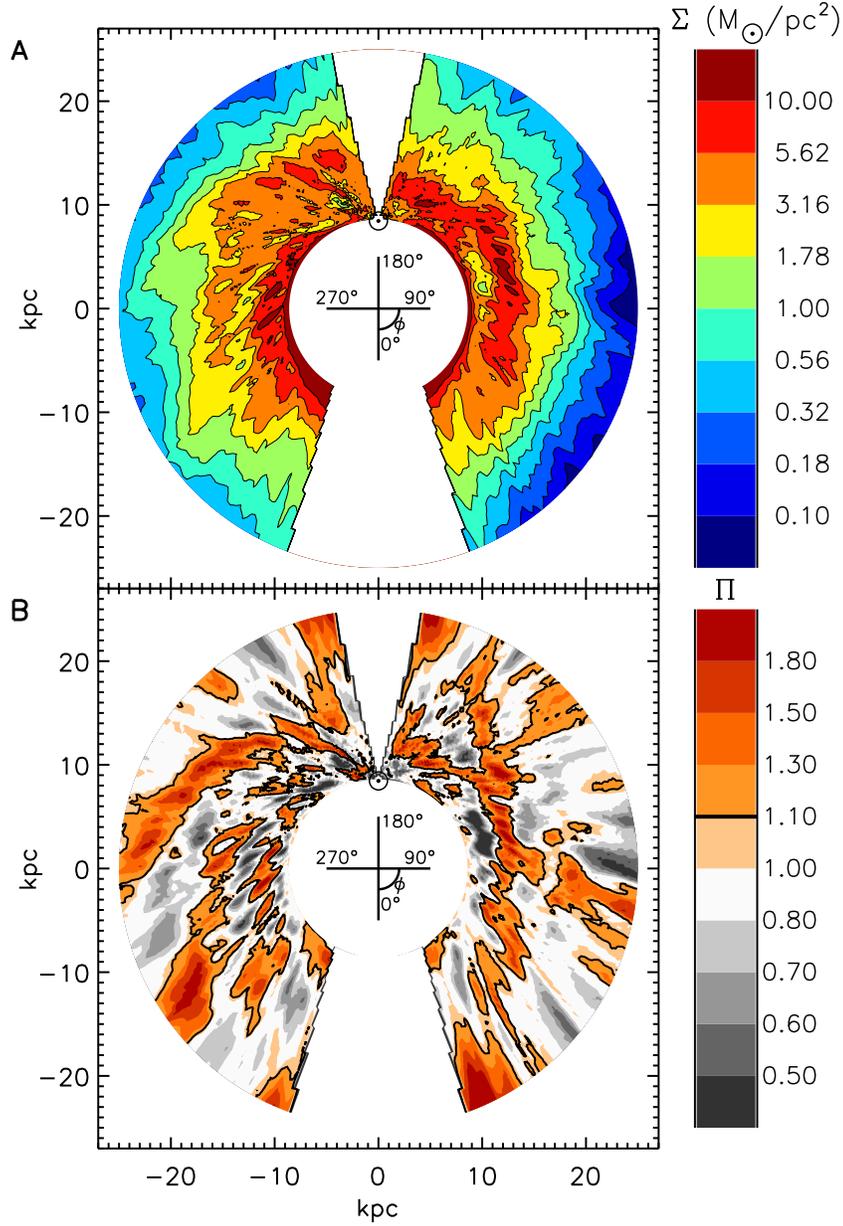}%
\caption{
A: Contour plot of the surface density $\Sigma(R,\phi)$. The location of the Sun is marked at (0, 8.5 kpc) by the solar symbol $\odot$. The regions excluded due to unreliable distances are the large blank wedges near the Sun-Galactic center line.  B: Contour plot of $\Pi(R,\phi)$, as defined in the text. Colored regions are overdense compared to the local median, whereas grayscale regions are underdense. The solid contour  marks the line $\Pi=1.1$. The values of $\Pi$ for the different contour levels are given by the colorbar.}
\end{center}
\end{figure}

 Values of surface density ratios in Fig.~1 range from a minimum of $\Pi=0.13$  to a maximum of 10.36. The vast majority of points have values in the range 0.6 to 1.8, implying that the arm-to-interarm surface density ratio is about 3.
 The typical value of $\Pi$ does not vary strongly as a function of $R$ along an arm, or even from arm to arm. 
 The $\Pi=1$ line does not  signify the border of a spiral arm, so Fig.~1 does not provide a meaningful description of the width of the arms. Furthermore, the intrinsic velocity dispersion serves to widen the apparent arms, especially at large radii. 
 
 Several long spiral arms appear clearly on the map, but the overall structure does not have the  reflection through the origin symmetry of a  ``grand-design'' spiral. Easily identified spiral arms cover a larger area in the south than in the north; there are three arms in the southern half of the diagram. Calculating $\eta$ over a smaller range of $R$ and $\phi$ reveals more structure in the arms, but does not change their positions. 
Each of these arms has already been detected in previous maps out to about 17 kpc for the two arms closer to the anticenter \cite{HJK1982} and out to 24 kpc for the `outer' arm\cite{MDGG2004}.
The modified unsharp masking technique allows us to trace the arms further; these arms run coherently over a length of nearly 30 kpc in the outer Galaxy alone. Near the Sun, the overdense region in Fig.~1 at $R\approx10$ kpc and $\phi\approx160^\circ$ is associated with the Perseus arm; however, there is a  well-known discrepancy between photometrically
and kinematically determined distances  in this region \cite{M1968, XRZM2006}. The well-defined underdense interarm regions in the south are also quite striking. There are clear underdense regions between all three of the arms in the south. In the north, a strongly underdense region is apparent inside the arm at $R\approx13$ kpc.
 
 When using the Doppler shift velocity of an emission line to find the distance to the emitting gas, small-scale velocity perturbations result in distortions of the distance measure because the assumed rotation structure is inaccurate.
In fact, peaks in 21 cm emission intensity are more likely to result from a combination of density and velocity perturbations \cite{B1971,TL1972} than solely from HI overdensities. 
Physically, this occurs because a peak can be explained by a local density maximum or by nearby gas flowing into or out of that location. 
If velocity perturbations result from spiral structure as described in density wave theory \cite{Y1969}, then Fig.~1 will not represent the true perturbed surface density but will still be a reasonable representation of the spiral pattern \cite{HJK1982}. A dynamical model that self-consistently incorporates density and velocity perturbations is required to compensate for this effect, but is beyond the scope of this paper.

We also map the thickness of the HI gas as a function of position by calculating the second moment, $T$, of the density distribution $\rho$ for each point in $(R,\phi)$.
When we perform the modified unsharp masking on $T(R,\phi)$ to create $\Gamma(R,\phi)$, spiral structure is evident (Fig.~2). The quantity $\Gamma$ is a direct equivalent to $\Pi$ as defined in Eqn.~1. The value of $\Gamma$ ranges in magnitude from 0.32 to 3.29, although the dynamic range of the majority of the points lie within 0.55 to 1.5.
We plot the $\Pi=1.1$ overdensity contour from the surface density perturbation map (Fig.~1) on top of the thickness perturbation map (Fig.~2), showing that
there is a good match between the arm positions as calculated from the surface density and the thickness; the thickness of the HI layer is smaller in the arms than in the rest of the disk.  The correlation remains good even if we do not apply a thickness filter to the data, or if we use another definition of the thickness \cite{HJK1982,LBH2006a}. The two maps are not  independent, however; changing the density distribution in the arms will necessarily produce changes in the measured thickness unless the perturbations are distributed in the same way as the initial distribution.
The alignment of overdensities with regions of reduced thickness was suggested previously \cite{HJK1982}. This alignment has not been observed in other galaxies because surface density maps are most easily made for face-on galaxies where there is no information about the thickness of the gas layer.
 
\begin{figure}
\begin{center}
\includegraphics[angle=90,scale=.7]{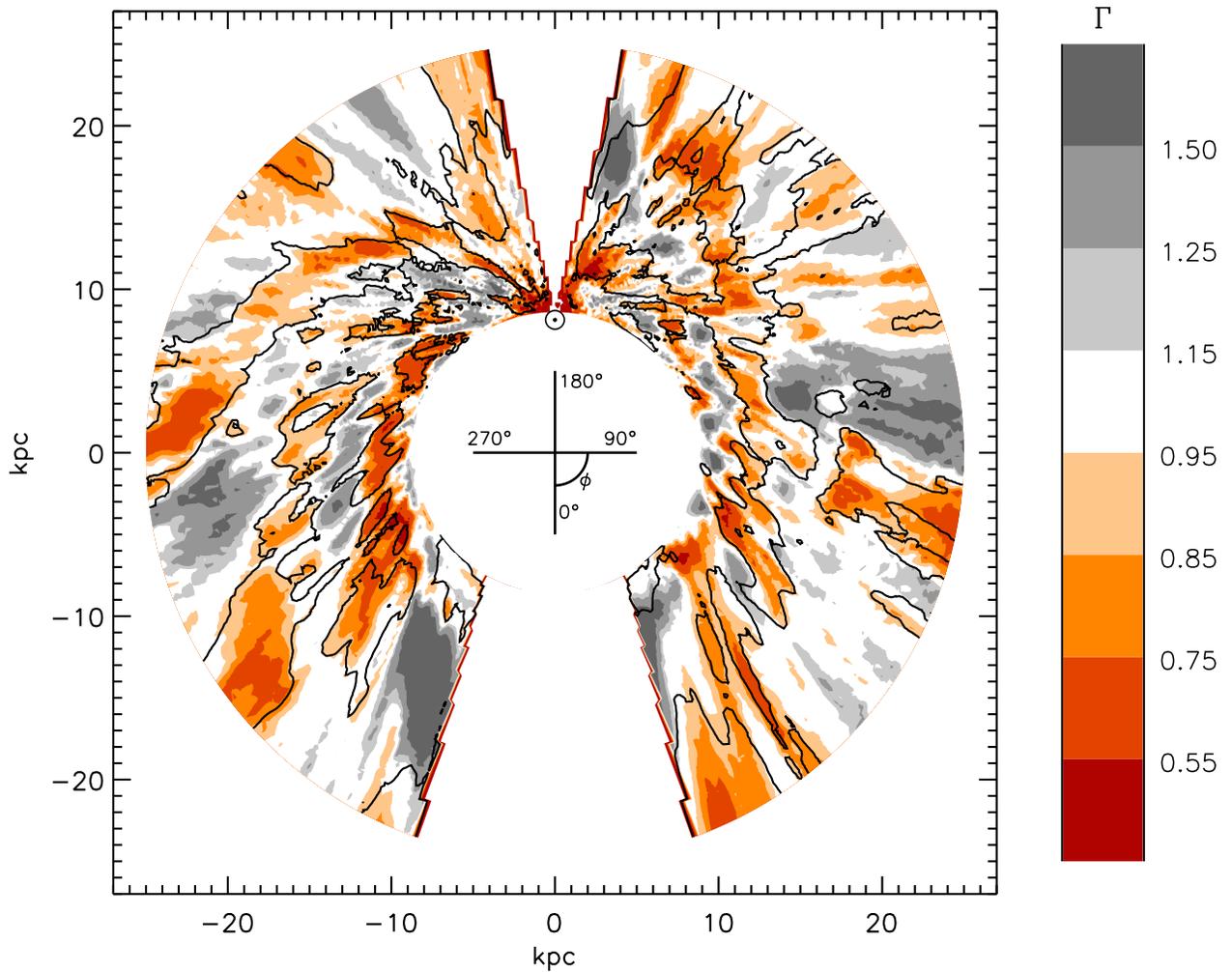}
\caption{
A contour plot of the perturbations in the gas thickness, $\Gamma(R,\phi)$. Colored regions have reduced thickness compared to the local median, whereas grayscale regions have larger thicknesses. The perturbation levels for the different contour levels are given by the colorbar. The solid contour  marks the line $\Pi=1.1$, the same as in Fig.~1, to show the alignment of the overdense surface densities with the thinner gas regions.  }\end{center}
\end{figure}

The radial profile of the HI disk has been a matter of controversy for many years. A sharp falloff in HI emission as a function of velocity has long been known \cite{D1971}, but this need not correspond to an abrupt  radial cutoff in the disk density \cite{KTG1978}. Velocity dispersion will cause features to be smeared along the line-of-sight by confusing the velocity-distance transformation, resulting the radially elongated features near the edges of maps (Fig.~3).
The radial extent of the spiral arms provides a minimum cutoff radius for the Galactic gas disk; in other words, it is not possible for the gas to have spiral structure beyond where the HI disk ends. This radius is only a lower limit because it is possible that there is gas beyond where the spiral structure ends that does not participate in the spiral structure, or that past some radius the arms are too weak to be detected by the unsharp masking.  Both the surface density and thickness perturbation maps (Fig.~3) change from spiral patterns to features elongated along the line-of-sight  near 25 kpc Galactocentric radius. This is most clearly seen in the south; the transition radius is not immediately obvious in the north. Thus, the HI gas disk must extend to at least 25 kpc from the Galactic center in the south, about three times the Sun-Galactic center distance. A related conclusion is that gas within the cutoff radius is kinematically settled into a disk; otherwise it would be unlikely to respond to the spiral density waves. 

\begin{figure}
\begin{center}
\includegraphics[angle=0,scale=.7]{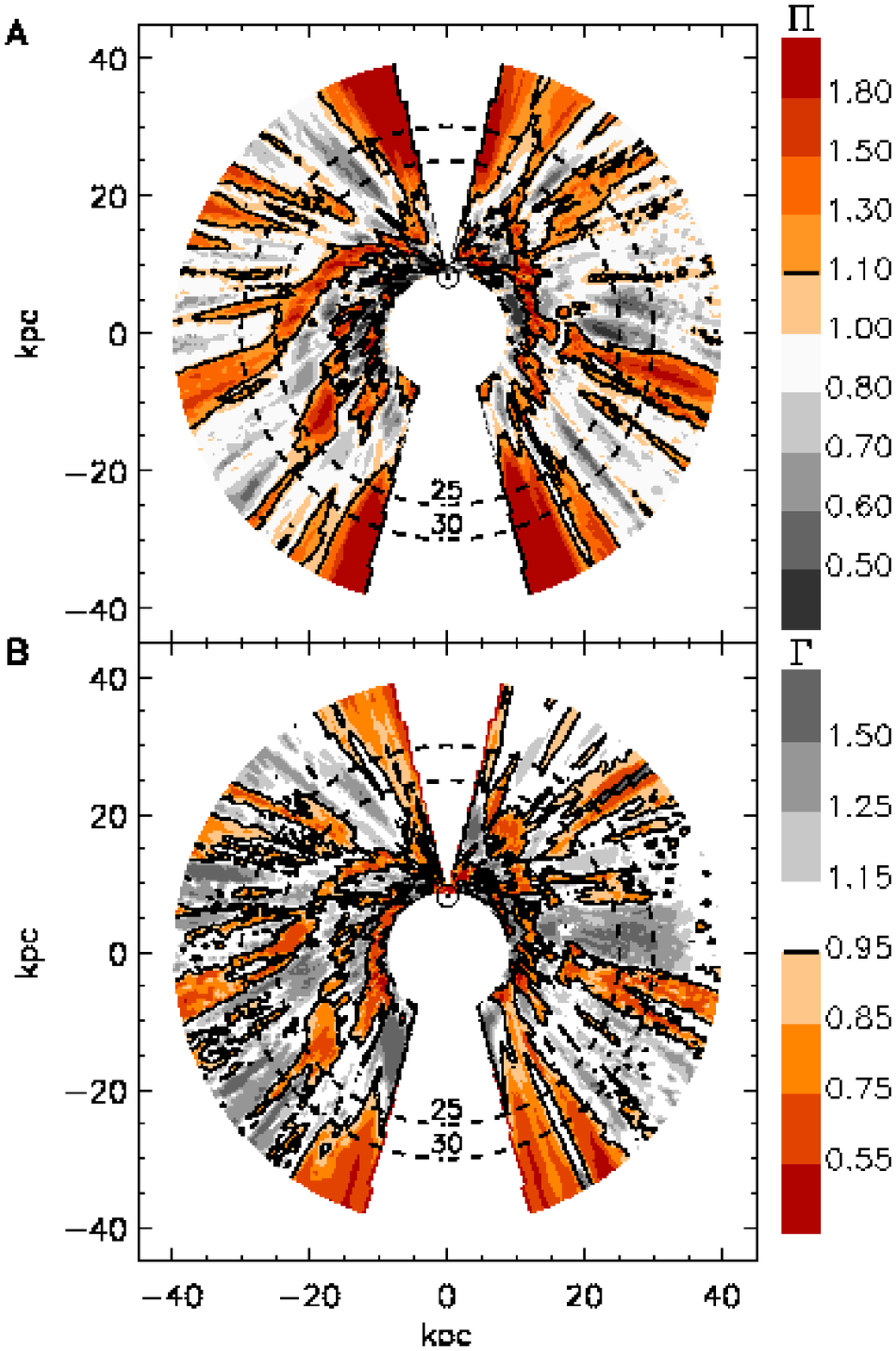}
\caption{
A: A contour map of the modified unsharp masked surface density, $\Pi(R,\phi)$, out to 40 kpc. B: A contour map of the modified unsharp masked disk thickness, $\Gamma(R,\phi)$. The dashed circles are lines of constant Galactocentric radius, marked in kpc.}
\end{center}
\end{figure}

It is useful to fit four-armed models to our density perturbation map. We use logarithmic spiral arms that start at the Solar circle:
\begin{equation}
\log(R/R_0)=(\phi(R)-\phi_0)\tan\psi
\end{equation}
where  $\psi$ is the pitch angle  and $\phi_0$ is the Galactocentric azimuth at the Solar circle. Our fitting method is designed to trace the regions of gas overdensity.
For each of the four arms  apparent in Fig.~1, we investigate an evenly spaced grid of these two free parameters for ranges of values that connect the overdense contours. 
 For each combination of $\psi$ and $\phi_0$, we linearly interpolate the value of $\Pi$ for the locus of points along each arm.  Any points that fall in the excluded regions are ignored. We use the median of the list of interpolated values as a measure of the goodness of fit for each curve. In this scheme, arms with values of $\psi$ and $\phi_0$ that trace overdense regions will naturally have a large median, and thus a large goodness-of-fit. 
The best-fit values of $\psi$ and $\phi_0$ for each of the four arms  are given (Table 1). Other fits that connect a different set of features in the map could be drawn because assigning a unique arm pattern to a map is  not possible.  We find pitch angles for the outer arms in the range 20$^\circ$-25$^\circ$; this is larger than the  value of $\psi\approx13^\circ$ averaged over a variety of tracers \cite{V2005}. This does not necessarily imply a disagreement, however, as the arms could be unwinding in their outer regions.

Various  models of the locations of the arms have been proposed. We compare our map to a model derived from regions of ionized hydrogren \cite{MWC1953,GG1976,WCVWS1992}; the model consists of two pairs of mirror symmetric arms  following logarithmic spirals. We denote this as  the symmetric model (Fig.~4).
  The symmetric model fits $\Pi$ reasonably well over much of  the southern sky; the agreement is poor in the north where the spiral structure is less prominent, possibly becayse of the larger thickness of the northern gas \cite{LBH2006a}. Gas that is  dynamically warmer is less likely to respond to spiral density waves, and the azimuthally averaged thickness of the northern gas is nearly twice that of the southern gas at $R=20$ kpc. 

\begin{figure}
\begin{center}
\includegraphics[angle=0,scale=.65]{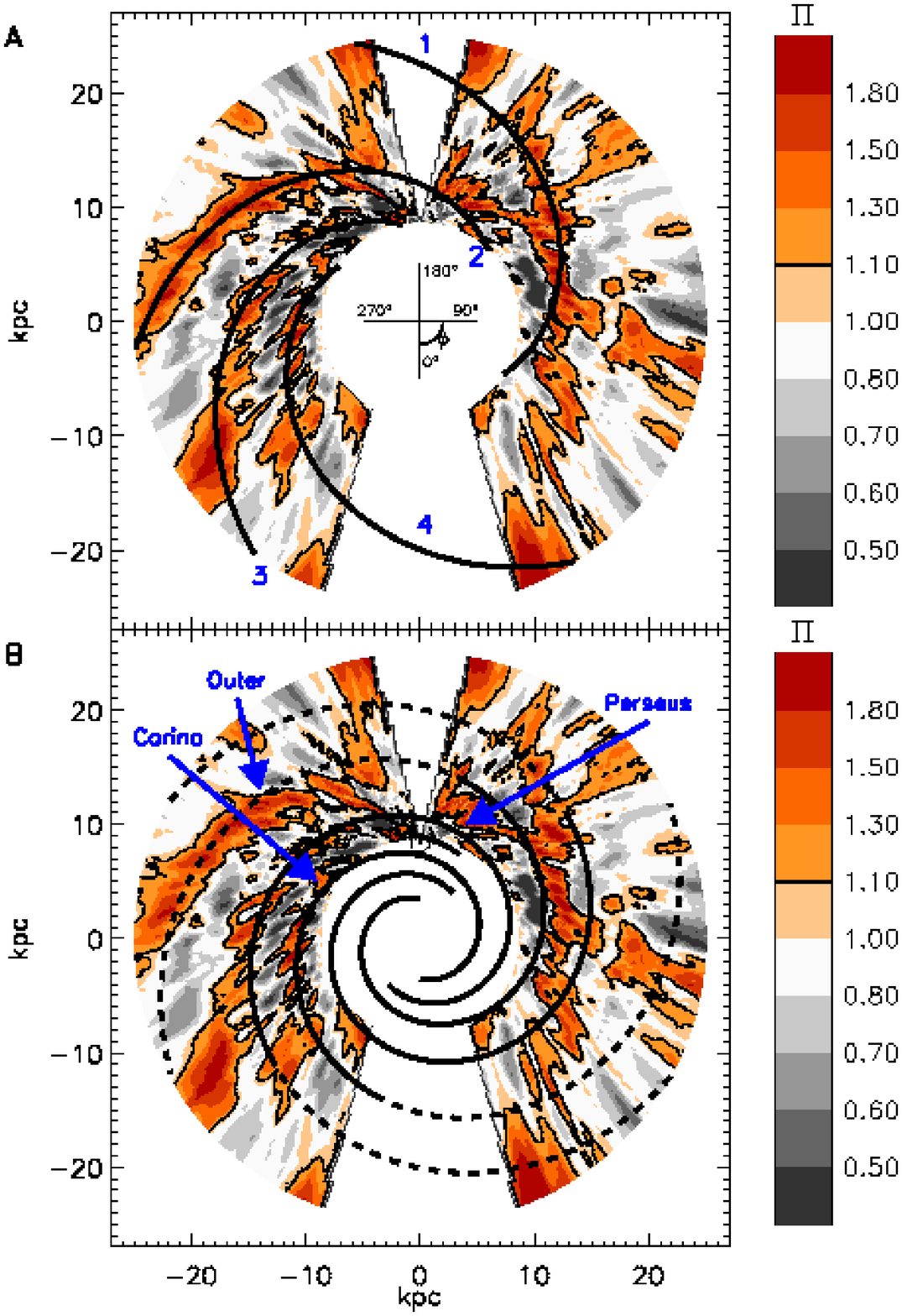}
\caption{
A:  The same contour plot as in Fig.~1, with a four armed logarithmic spiral fit overplotted. The fitting method is described in the text. Other fits that connect different features are possible. B: The same contour plot as in Fig.~1, with the four armed symmetric spiral model overplotted \protect\cite{WCVWS1992}. The solid lines represent the model over its claimed range of validity; the dashed lines are an extension beyond that range. The unlabeled short line near the Sun is the local Orion arm. The width of the model arms is arbitrary. 
}\end{center}
\end{figure}

\begin{table}
\begin{center}
\begin{tabular}{cccc}
Arm & $\psi$ (deg) & $\phi_0$ (deg) \\ \hline
1&24&56\\
2 & 24 & 135 \\
3 & 25 & 189 \\
4& 21 & 234 \\
\end{tabular}\caption{Fits to pitch angles $\psi$ and angles at which the arms cross the Solar circle $\phi_0$. The arm numbers correspond to Fig.~4A. }
\end{center}
\end{table}

There are several places where the symmetric model deviates from the data. 
For example, the arm in the north ($R\approx 13$ kpc)  falls in between two of the model's  arms; forcing the arms to be mirror imaged pairs is too strong a restriction.  Features near the excluded regions could result from large scale ordered velocity structure that has not been included in our rotation model.  Elliptical streamlines with $m=2$ could cause such an effect \cite{LBH2006a}. Images of other galaxies suggest that the spiral arms may bifurcate into spurs in the outer disk.  The structure of the Perseus and Carina arms past $R\approx20$ kpc is suggestive of this behavior.

\bibliographystyle{science.bst}


\begin{scilastnote}
\item We would like to thank Peter Kalberla for providing a copy of the LAB HI survey. Josh Peek, Conor Laver, and Tim Robishaw gave helpful advice regarding plots. ESL and LB are supported by NSF grant AST 02-28963. CH is supported by NSF grant AST 04-06987.
\end{scilastnote}

\end{document}